\documentclass[
journal=prl,
doi=true, superscriptaddress,
manuscript=article,layout=twocolumn, reprint, floatfix
]{revtex4-1}

\usepackage[version=3]{mhchem} %

\usepackage{dcolumn}%
\usepackage{bm}%
\usepackage{amsfonts}
\usepackage{bbm}
\usepackage{todonotes}
\usepackage{tikz}
\usepackage{cool}
\usepackage{dirtytalk}
\usepackage{physics}
\usepackage[fleqn]{mathtools}
\usepackage{isomath}
\usepackage{lmodern}
\usepackage[english]{babel}
\usepackage{multirow} 
\usepackage{pdfpages}
\usepackage{pgffor}

\makeatletter
\AtBeginDocument{\let\LS@rot\@undefined}
\makeatother

\makeatletter
\def\bbl@set@language#1{%
  \edef\languagename{%
    \ifnum\escapechar=\expandafter`\string#1\@empty
    \else\string#1\@empty\fi}%
  \@ifundefined{babel@language@alias@\languagename}{}{%
    \edef\languagename{\@nameuse{babel@language@alias@\languagename}}%
  }%
  \select@language{\languagename}%
  \expandafter\ifx\csname date\languagename\endcsname\relax\else
    \if@filesw
      \protected@write\@auxout{}{\string\select@language{\languagename}}%
      \bbl@for\bbl@tempa\BabelContentsFiles{%
        \addtocontents{\bbl@tempa}{\xstring\select@language{\languagename}}}%
      \bbl@usehooks{write}{}%
    \fi
  \fi}
\newcommand{\DeclareLanguageAlias}[2]{%
  \global\@namedef{babel@language@alias@#1}{#2}%
}
\makeatother

\DeclareLanguageAlias{en}{english}

\begin{document}

\title{Inexpensive modelling of quantum dynamics \\
using path integral generalized Langevin equation thermostats}

\author{Venkat Kapil}
\email{venkat.kapil@epfl.ch}
 \affiliation{Laboratory of Computational Science and Modelling, Institute of Materials, Ecole Polytechnique F\'ed\'erale de Lausanne, Lausanne 1015, Switzerland}
 
\author{David M. Wilkins}
\affiliation{Laboratory of Computational Science and Modelling, Institute of Materials, Ecole Polytechnique F\'ed\'erale de Lausanne, Lausanne 1015, Switzerland}

\author{Jinggang Lan}
\affiliation{Department of Chemistry, University of Z\"urich, Z\"urich, Switzerland}

\author{Michele Ceriotti}
\affiliation{Laboratory of Computational Science and Modelling, Institute of Materials, Ecole Polytechnique F\'ed\'erale de Lausanne, Lausanne 1015, Switzerland}%

\begin{abstract}
The properties of molecules and materials containing light nuclei are affected by their quantum mechanical nature. Modelling these quantum nuclear effects accurately requires computationally demanding path integral techniques. Considerable success has been achieved in reducing the cost of such simulations by using generalized Langevin dynamics to induce frequency-dependent fluctuations. 
Path integral generalized Langevin equation methods, however, have this far been limited to the study of static, thermodynamic properties due to the large perturbation to the system's dynamics induced by the aggressive thermostatting.
Here we introduce a post-processing scheme, based on analytical estimates of the dynamical perturbation induced by the generalized Langevin dynamics, that makes it possible to recover meaningful time correlation properties from a thermostatted trajectory. We show that this approach yields spectroscopic observables for model and realistic systems which have an accuracy comparable to much more demanding approximate quantum dynamics techniques based on full path integral simulations.

\end{abstract}

\maketitle

Vibrational spectroscopic techniques -- from conventional infrared (IR) and Raman to advanced femtosecond pump-probe~\cite{perakis2016vibrational,woutersen_femtosecond_1997}, sum-frequency generation, second harmonic scattering~\cite{roke_nonlinear_2012, shen_optical_1989}, and multi-dimensional vibrational spectroscopy~\cite{bakker2009vibrational} -- are a cornerstone of chemistry~\cite{perakis_vibrational_2016}. These techniques have a multitude of applications such as the characterization of functional groups in chemical systems~\cite{noauthor_map:_2017}, the determination of the atomistic mechanisms of phase transitions through insight into chemical environments~\cite{xue_detection_2013}, and identification of unique structural fingerprints of molecular crystals~\cite{han_predicting_2019}. The use of atomistic simulations for the computation of spectroscopic properties facilitates the interpretation of these experiments and provides support to the  characterization of novel materials. 

Even neglecting effects that go beyond the Born-Oppenheimer (BO) decoupling of electronic and nuclear degrees of freedom, accurate calculations of the vibrational spectra of materials require an explicit treatment of the quantum dynamics of the nuclear degrees of freedom~\cite{chandler_exploiting_1981} on the electronic ground state potential energy surface. Quantum dynamics is in principle exactly obtained from the solution of the time dependent Schr\"odinger equation for the nuclei, but this is only practical for systems containing a handful of degrees of freedom~\cite{li_vibrational_2001, vendrell_full_2007}. Condensed phase systems can be studied~\cite{habershon_ring-polymer_2013} either by an exact treatment of the quantum dynamics of a subset of the nuclear degrees of freedom~\cite{wang_ab_2011}, or through classical dynamics on the quantum free energy surface of the nuclei~\cite{hele_boltzmann-conserving_2015, miller_semiclassical_2001}. Among the methods in the latter class, several of the most popular ones are based on the imaginary time path integral framework -- such as (thermostatted) ring polymer molecular dynamics~\cite{craig_quantum_2004, rossi_how_2014} ((T)RPMD), centroid molecular dynamics~\cite{cao_formulation_1994, hone_comparative_2006} (CMD) and the recently developed quasi-centroid molecular dynamics~\cite{trenins_path-integral_2019} (QCMD).
These methods ignore real time coherence but include effects arising from equilibrium quantum fluctuations and have been validated on several model systems and small molecules for which exact or highly accurate results are available~\cite{hele_boltzmann-conserving_2015, rossi_how_2014, braams_short-time_2006, rossi_how_2014, hone_comparative_2006}. While these methods show great promise for accurate determination of spectroscopic properties~\cite{medders_representation_2015,litman_elucidating_2019, marsalek_quantum_2017}, their cost remains high when combined with a potential energy surface computed by \emph{ab initio} electronic structure methods. 

Among the many methods that have been introduced in the past decade to accelerate the convergence of path integral calculations~\cite{markland_nuclear_2018}, those that combine path integral molecular dynamics with a generalized Langevin equation~\cite{ceriotti_accelerating_2011,ceriotti_efficient_2012,brie+16jctc} can be applied transparently to empirical, machine learning or first principles simulations. 
They have been used to evaluate all sorts of thermodynamic properties, including structural observables~\cite{ceriotti_nuclear_2013}, free energies~\cite{liu_barely_2019}, momentum distributions~\cite{ceriotti_efficient_2012}, and quantum kinetic energies~\cite{wang_quantum_2014} with a reduction in computational effort varying between a factor of 5 at ambient conditions to a factor of 100 at cryogenic temperatures~\cite{uhl_accelerated_2016}.  The aggressive thermostatting used to impose quantum fluctuations, however, significantly disrupts the dynamics of the system, and common wisdom is that the calculation of dynamical properties using PIGLET is impossible. 

Here we present a simple post-processing strategy that makes it possible to reconstruct dynamical properties from trajectories generated using PIGLET, leading to a dramatic reduction in  the cost of including quantum effects in spectroscopic quantities. We show that when applied to model systems the accuracy of the scheme is on par with that of conventional path integral dynamical schemes, aside from a small residual broadening of the spectra. 
We then demonstrate the usefulness our approach by computing the IR and Raman spectra of solid and liquid water,
using state-of-the-art machine-learning interatomic potentials, dipole moment, and polarizability surfaces. 

The time evolution of a one-dimensional harmonic oscillator with unit mass, position $q$, momentum $p$ and frequency $\omega_0$ subject to a generalized Langevin equation can be expressed in the Markovian form
\begin{equation}
\begin{split}
  \dot{q}=&p\\
\!\left(\! \begin{array}{c}\dot{p}\\ \dot{\mathbf{s}} \end{array}\!\right)\!=&
\left(\!\begin{array}{c}-\omega_0 q\\ \mathbf{0}\end{array}\!\!\right)
\!-\!\left(\!
\begin{array}{cc}
a_{pp} & \mathbf{a}_p^T \\ 
\bar{\mathbf{a}}_p & \mathbf{A}
\end{array}\!\right)\!
\left(\!\begin{array}{c}p\\ \mathbf{s}\end{array}\!\right)\!+\!
\left(\!
\begin{array}{cc}
b_{pp} & \mathbf{b}_p^T \\ 
\bar{\mathbf{b}}_p & \mathbf{B}
\end{array}\!\right)\!
\left(\!\begin{array}{c}\multirow{2}{*}{$\boldsymbol{\xi}$}\\ \\\end{array}\!\right).
\end{split}
\label{eq:Markov}
\end{equation}
following the notation of Ref.~\cite{ceriotti_colored-noise_2010}. Here, $\boldsymbol{\xi}$ is a vector of uncorrelated Gaussian random numbers, $\mathbf{s}$ is a vector of auxiliary momenta, and $\mathbf{A}_{\square}$ and $\mathbf{B}_{\square}$ respectively define the so called drift and diffusion matrices of the GLE. The power spectrum of the time correlation function of a linear operator, such as the velocity, is known to be~\cite{rossi_fine_2017},
\begin{equation}
    \mathcal{C}_{pp}(\omega, \omega_0) = \frac{1}{\left[\mathbf{C}_{qp}(\omega_0)\right]_{pp}}  \left[\frac{\mathbf{A}_{qp}(\omega_0)}{\mathbf{A}_{qp}(\omega_0) + \omega_0^2} \right]_{pp}.
\label{eq:cpp}
\end{equation}
where $\mathbf{C}_{qp}$ is the static correlation matrix readily obtained from $\mathbf{A}_{qp}$ and $\mathbf{B}_{qp}$~\cite{ceriotti_colored-noise_2010}. 
If one approximates the response of a system to be described by an ensemble of oscillators with a microcanonical density of states $g(\omega)$ and oscillator strength $\iota(\omega)$, the spectrum can be written as the convolution of the weighted density of states with the kernel of Eq.~\eqref{eq:cpp}:
\begin{equation}
    C^{\textrm{GLE}}(\omega) =  \int \textrm{d}\omega' ~g(\omega') \iota({\omega'}) ~\mathcal{C}_{pp}(\omega, \omega'). \label{eq:convolution}
\end{equation}
This result follows by the fact that independent GLEs applied in the Cartesian basis act as if they acted in the normal modes basis, and so each vibrational mode can be treated independently~\cite{ceriotti_langevin_2009}. 
Similar results can be obtained by linking expressions based on the Fourier transform of the dielectric responses of the system to the velocity correlation function, e.g. for the dipole derivative correlation spectrum
\begin{equation}
C_{\dot{\mu}\dot{\mu}} (t) = \left<\frac{\partial{\mu}(0)}{\partial{x}} v(0)v(t) \frac{ \partial{\mu}(t)}{\partial{x}}  \right>  \approx
\left<\left|\frac{\partial{\mu}}{\partial{x}}\right|^2\right> \left<v(0)v(t)\right>.
\label{eq:mumu}
\end{equation}
The Fourier transform of the velocity correlation function in the presence of the GLE can be obtained using a similar convolution analogous to Eq.~\eqref{eq:convolution}.
Note that the second step in Eq.~\eqref{eq:mumu} involves the assumption of a linear dipole operator. The corrections associated with a non-linear operator could be computed analytically, but cannot be disentangled from the linear spectrum. We will verify that non-linearities in the dielectric response do not introduce major artefacts by computing spectra using machine-learned dipole and polarizability surfaces.

As discussed in Ref.~\citenum{rossi_fine_2017}, for a classical trajectory $C^{\textrm{GLE}}(\omega)$ can be deconvoluted to obtain the underlying micro-canonical density of states by solving the inverse problem using $\mathcal{C}_{pp}(\omega, \omega')$ as the convolution kernel. 
In the same work, it was shown that these relations describe accurately the dynamics of strongly anharmonic systems such as liquids, despite having been derived in the harmonic limit, and that they can be applied to correct time correlation functions not only for classical Langevin dynamics, but also %
in the case of the ``quantum thermostat" -- a GLE thermostat that enforces approximate quantum fluctuations with a single MD trajectory. 
While the results of the quantum thermostat  display qualitative features associated with nuclear quantum effects such as the red shift of the stretching and bending modes, and the increase in the diffusion coefficient, they suffer from the fact that the quantum thermostat is not systematically improvable, and offers only a qualitative description of the quantum Boltzmann distribution.

In order to overcome these limitations, we show that a deconvolution scheme can also be applied to the case of spectra obtained with the PIGLET technique. 
Given that in PIGLET simulations the centroid of the ring polymers is subject to a classical Boltzmann sampling termostat,  we correct the velocity autocorrelation function of the centroid using the deconvolution kernel defined by the thermostat that acts on it, assuming that the (non-equilibrium) GLE thermostats that act on the internal modes of the ring-polymer do not affect the dynamics of the centroid. 
This assumption is exact in the harmonic limit for any number of beads, as the time evolution of the centroid is decoupled from the other  ring polymer modes. In 
the limit of large number of beads  PIGLET tends to standard canonical sampling, and the (purified) centroid dynamics should converge to that of TRPMD.

\begin{figure*}[tbh]
\begin{center}
\includegraphics[width=0.45\textwidth]{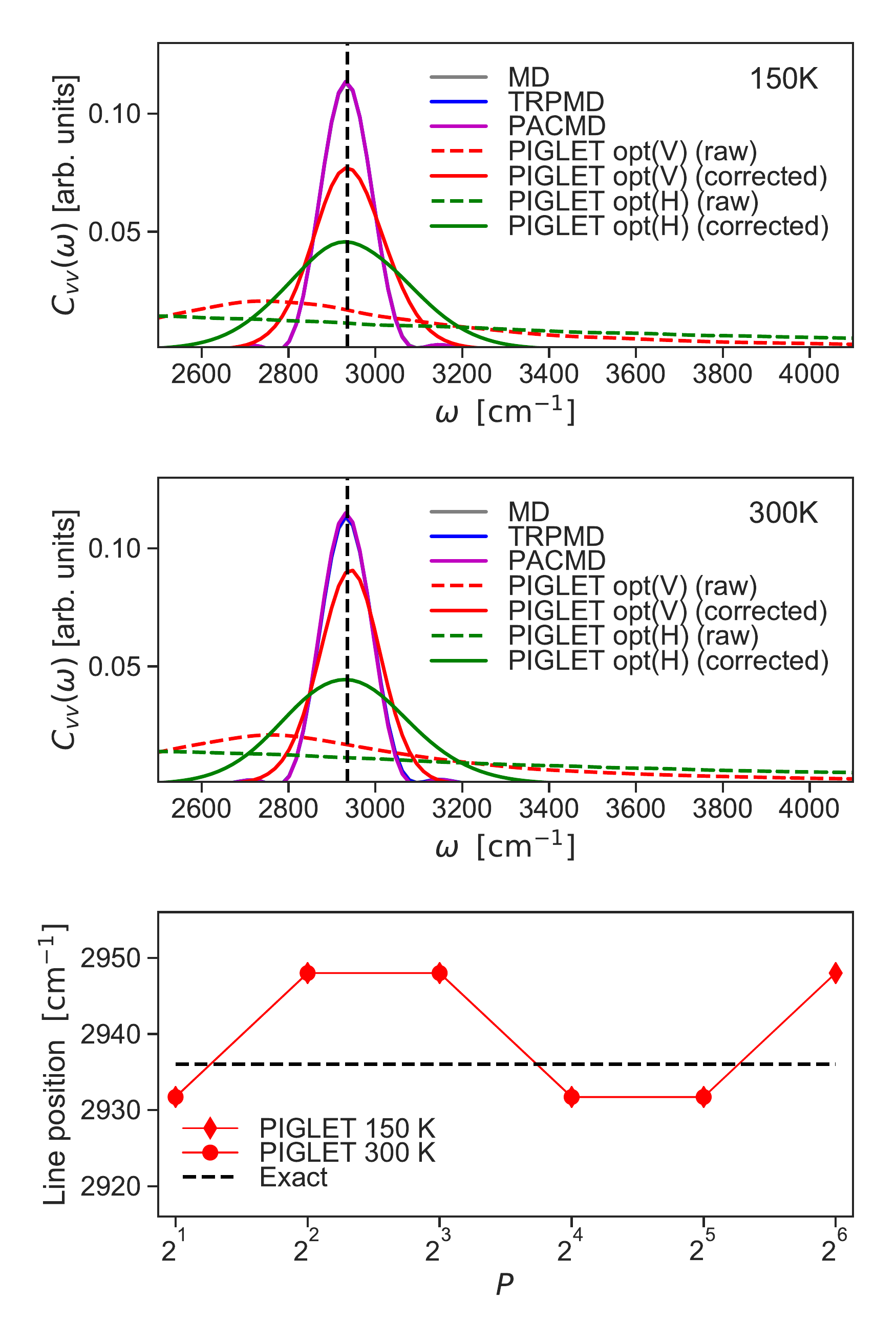}\ \ \ \ 
\includegraphics[width=0.45\textwidth]{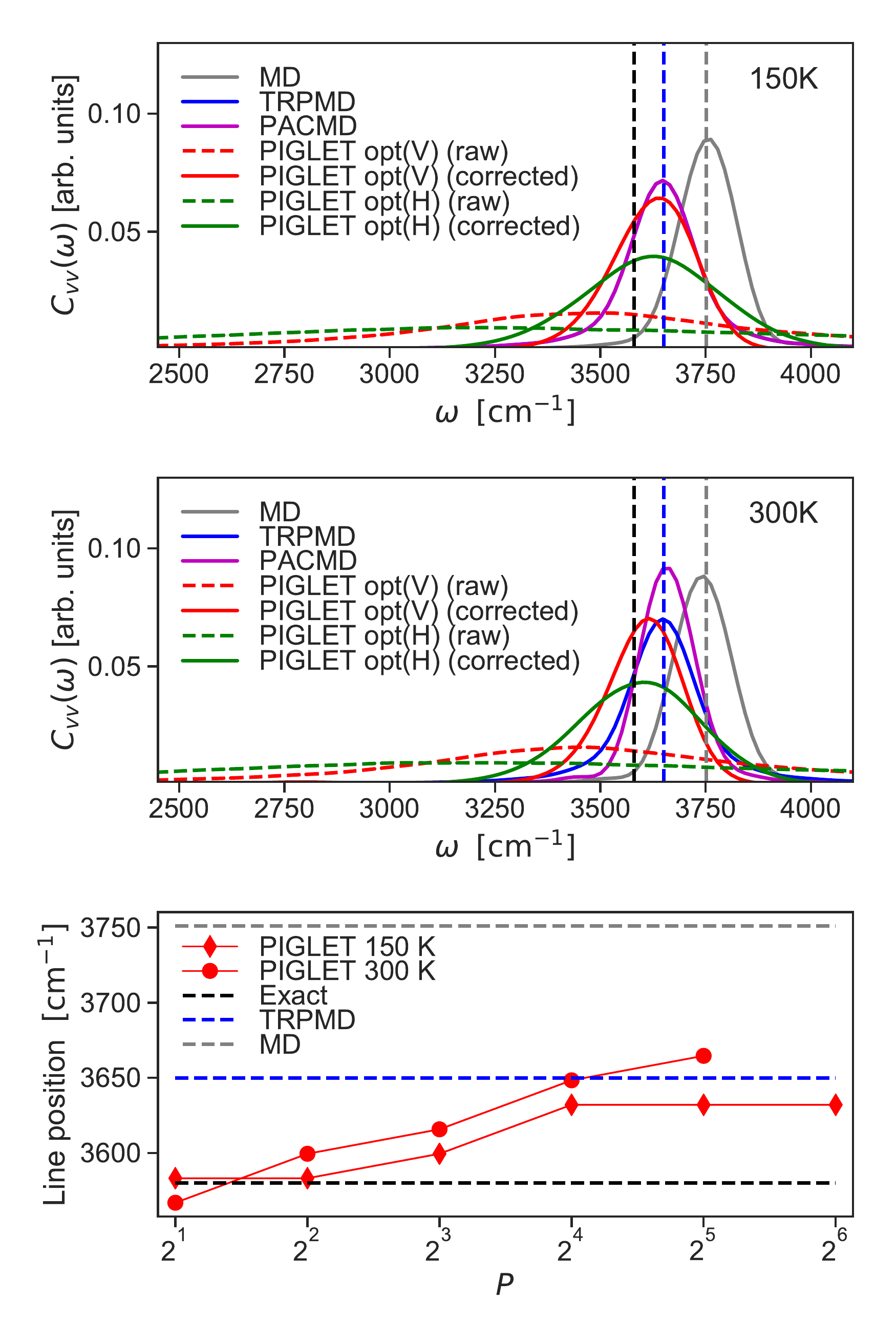}
\end{center}
\caption{Vibrational density of states as calculated by the velocity autocorrelation $C_{{v}{v}}(\omega)$ for a harmonic oscillator parametrized to model a C--H bond (left) and a Morse oscillator parametrized to model an O--H bond (right). Simulations were performed at 150\,K (top) and 300\,K (middle) using MD (grey), TRPMD (blue), PA-CMD (purple), and using the opt(H) (green) and opt(V) (red) variants of the PIGLET thermostat. The dashed and solid line show the spectra obtained using the PIGLET thermostat, and the corrected spectra using the deconvolution scheme described in this work, respectively. The bottom panel shows the converge of the line position using the opt(V) variant of PIGLET as a function of number of replicas $P$ at 150\,K (diamond markers) and 300\,K (circular markers). The dashed black line shows the $0\to1$ transition frequency.
\label{fig:1D}}
\end{figure*}

We begin illustrating the merits and the limits of this approach by studying the VDOS of simple and analytic model systems for which exact results are known. 
We consider a one-dimensional harmonic ${\small \hat{H} = \frac{\hat{p}^2}{2\mu} + \frac{1}{2} K \left(\hat{q}-q_0\right)^2,}$ with $\mu = 1694.9533$, $q_0=2.0598$ and $K = 0.3035~\text{a.u.}$ tuned to reproduce a typical C--H stretching vibration~\cite{witt_applicability_2009}.
In the harmonic limit methods such as TRPMD, CMD and even classical MD are known to deliver exact line positions. 
We have used two variants of the PIGLET thermostat -- that we will refer to as opt(V) and opt(H) -- with different parameterizations of the centroid thermostat. 
The parameters of the opt(V) and opt(H) thermostats have been tailored, respectively, to minimize the autocorrelation time of the potential and the total energy, and  the latter is roughly speaking twice as aggressive as the former~\cite{ceriotti_langevin_2009}.
As shown in the left panel of Fig. \ref{fig:1D}, the PIGLET thermostats have a detrimental effect on the VDOS, broadening and shifting the peaks to an extreme level, which is the reason why they are usually considered unsuitable for the calculation of dynamical properties.
The ``corrected spectra'' obtained by applying the deconvolution procedure, instead, have line positions that are in good agreement with the exact results at both 150\,K and 300\,K~\footnote{It can be shown that in the harmonic limit, for exact integration, results ought to be independent of temperature, so the agreement between the two temperatures indicate that the method is robust to integration and sampling errors.}. The line widths of the corrected spectra are slightly larger than those obtained with other methods, even after deconvolution.  
Between the two variants of PIGLET, opt(H) leads to higher Lorentzian broadening, which indicates that contrary to the case of a classical trajectory the deconvolution procedure cannot eliminate completely the effect of the thermostat, making the lineshape dependent on the parameterization of the GLE.
As shown in the bottom panel of Fig. \ref{fig:1D}, the line positions obtained by ``correcting" the opt(V) spectra agree with the exact results to within 15$\,\text{cm}^{-1}$, which is the frequency resolution of the computed spectra.

Encouraged by the observation that -- particularly for the weaker opt(V) centroid thermostat -- deconvoluted PIGLET spectra show harmonic vibrations at the correct frequency, and only a modest peak broadening,  we consider the case of a one-dimensional Morse oscillator. The parameters of the Hamiltonian ${\small \hat{H} = \frac{\hat{p}^2}{2\mu} + D \left(1 - e^{-\alpha \left(\hat{q}-q_0\right)}\right)^2}$ are tuned to describe the anharmonic O--H stretching mode in water~( $\mu=1741.0519$, $q_0=2.0598$ and $\alpha=1.1605~\text{a.u.}$, as in Ref.~\citenum{witt_applicability_2009}), which displays a large nuclear quantum effect in the frequency of the ${0 \to 1}$ transition~\cite{rossi_how_2014}. As shown in the top and middle panels of Fig. \ref{fig:1D}, the line positions obtained using all methods are blue-shifted with respect to the exact ${0 \to 1}$ transition frequency. Classical MD blue-shifts the line position by over 100$\,\text{cm}^{-1}$, while CMD and TRPMD, which both require 64 and 32 replicas at 150\,K and 300\,K, respectively, are shifted by around  60\,$\text{cm}^{-1}$. As shown in the bottom panel, the line positions of the deconvoluted PIGLET spectra remain close to the exact result for a small number of replicas and systematically converge to line positions of TRPMD as the number of replicas is increased. For the number of beads that typically converge structural observables to their quantum limit i.e. 16 replicas at 150\,K and 6 replicas at 300\,K~\cite{kapil_high_2016}, the results obtained using the proposed scheme are within 50\,$\text{cm}^{-1}$ of the exact result and in good agreement with TRPMD and CMD. As in the case of the harmonic oscillator the spectra obtained using the opt(H) variant of PIGLET are considerably broadened even after deconvolution. On the other hand, the spectra obtained using opt(V) are only slightly broader than those obtained with TRPMD, and give accurate line positions with a computational cost that is 3-4 times lower.

\begin{figure}[ht]
\centering
    \includegraphics[width=0.45\textwidth]{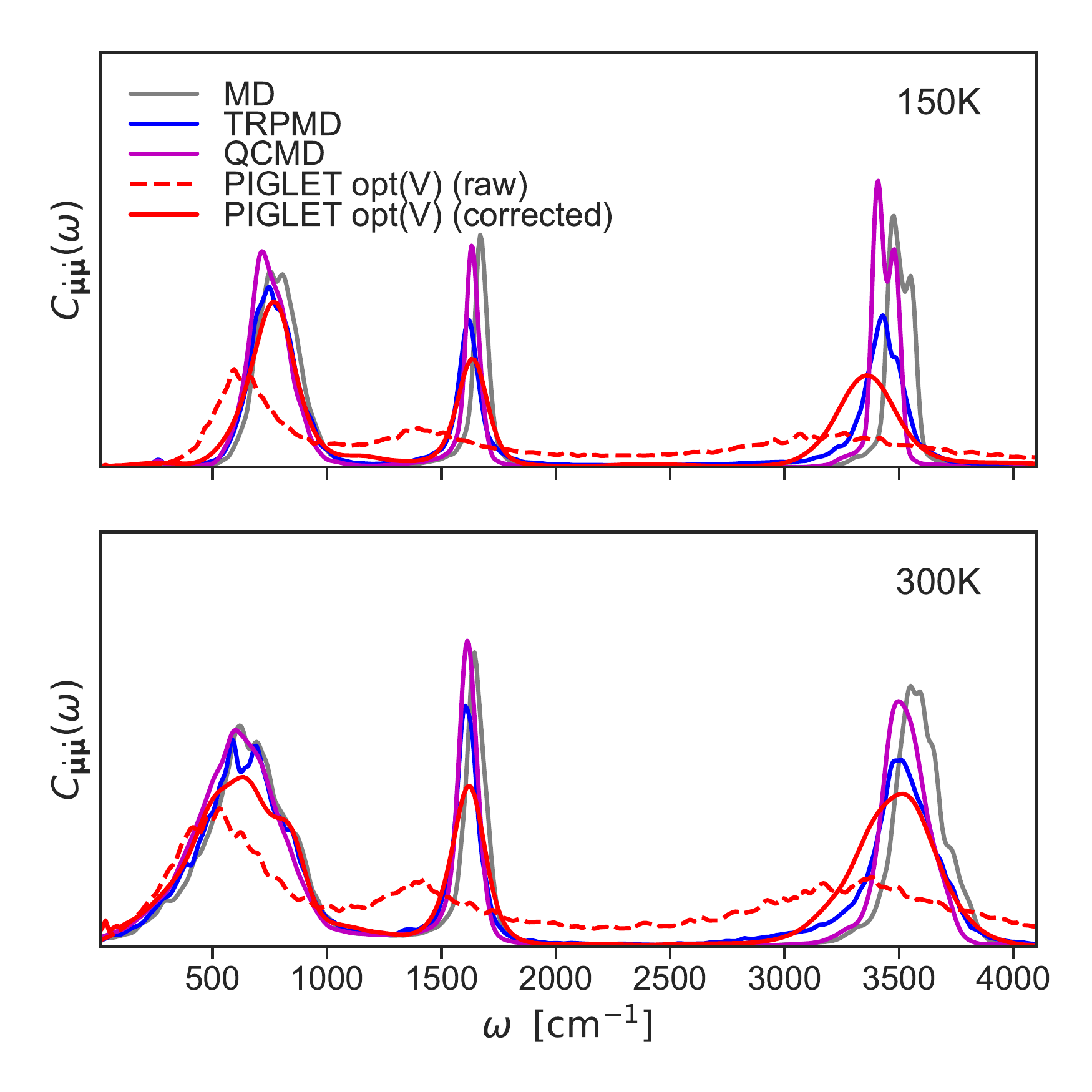}
\caption{Infra-red $C_{\dot{\bm{\mu}}\dot{\bm{\mu}}}(\omega)$ spectra as calculated by the autocorrelation of the time derivative of the dipole moment for hexagonal ice at 150\,K (top) and liquid water (bottom) at 300\,K, described with the q-TIP4P/f potential. Simulations have been performed using MD (grey), TRPMD (blue), QCMD (purple), and using opt(V) (red) variant of the PIGLET thermostat. The dashed and solid line show the spectra obtained using the PIGLET thermostat, and the corrected spectra using the deconvolution scheme described in this work, respectively. 
\label{fig:H2O}}
\end{figure}

\begin{figure}[ht]
\centering

\includegraphics[width=0.5\textwidth]{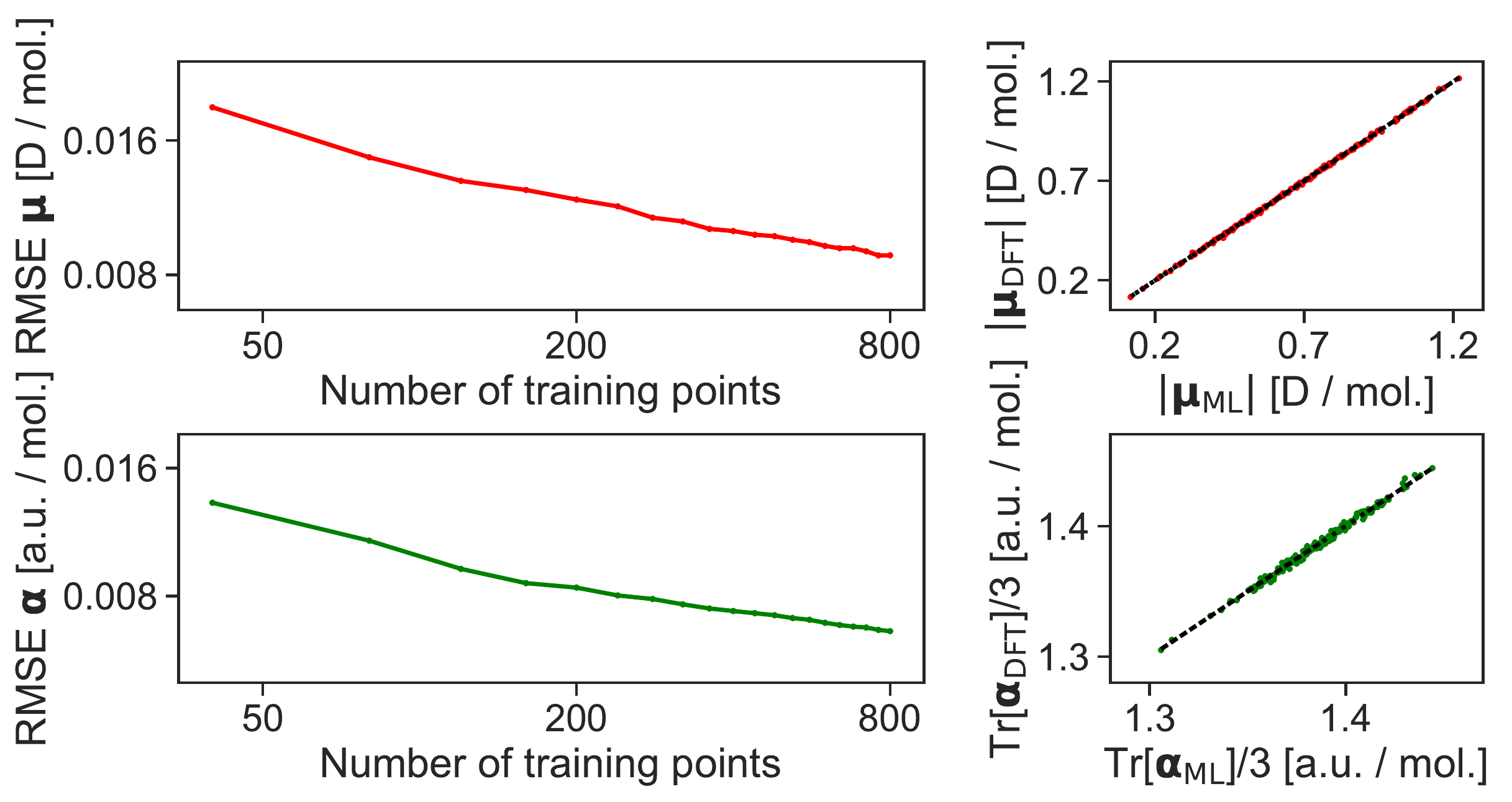}

\caption{Left: Learning curves for the $\mu$H$_{2}$O (top) and $\alpha$H$_{2}$O (bottom) models for the dipole moment $\boldsymbol{\mu}$ and polarizability $\boldsymbol{\alpha}$ of bulk water, showing the root mean square error (RMSE) in predicting $\boldsymbol{\mu}$ and $\boldsymbol{\alpha}$ of a testing set containing 200 frames as a function of the number of frames used to train the model.
Right: Parity plots (right) of the predictions of these models with ideal predictions shown in dashed black lines.
\label{fig:learning_curves}}
\end{figure}

Finally, we test the approach on condensed-phase systems. We begin by studying the IR spectrum of hexagonal ice at 150\,K and liquid water at 300\,K using the q-TIP4P/f water model~\cite{habershon_competing_2009} and a linear dipole moment surface, as used in a number of prior investigations~\cite{rossi_communication:_2014,trenins_path-integral_2019, willatt_approximating_2018}. The IR spectrum $C_{\dot{\bm{\mu}}\dot{\bm{\mu}}}(\omega)$ is calculated using the autocorrelation of the time derivative of the instantaneous dipole moment of the system ${\bm{\mu}}(t)$ and is normalized to integrate to unity. Since the opt(V) variant of PIGLET has been shown to consistently provide more resolved spectra, we report results only with this variant using 6 replicas at 300\,K and 16 replicas at 150\,K. Furthermore, we have not used PA-CMD as it is known to exhibit a curvature problem~\cite{ivanov_communications:_2010} at low temperatures. We instead present results from QCMD simulations taken from Ref.~\cite{trenins_path-integral_2019}.  As shown in the top and middle panels of Fig.~\ref{fig:H2O}, the thermostatted IR spectra are severely affected but show three distinct bands  that correspond to the librational, bending and stretching motion. All the bands are well resolved by applying the deconvolution procedure. 
At 300\,K, all three bands are in good agreement with TRPMD and CMD, aside from a slight broadening of the stretching band for the deconvoluted PIGLET spectra.
At 150\,K, instead, the librational and bending bands remain in good agreement with TRPMD and QCMD while the stretching band is broadened and red shifted by around 60 cm$^{-1}$ with respect to these methods. 
Given that in the case of gas phase water, for which exact quantum dynamical results are available, both QCMD and TRPMD yield consistently a blue-shift of 60\,$\text{cm}^{-1}$ for the stretching band, the discrepancy is comparable with the typical accuracy of much more demanding methods. 
After having shown the promise of our deconvolution scheme in recovering dynamical properties from GLE-accelerated spectra, we apply it to a more challenging problem, namely the prediction of IR and Raman spectra of water modelled with first-principles-quality machine-learning techniques.
In order to inexpensively evaluate the dipole and polarizability surfaces, 
we trained a new model based on the recently developed symmetry-adapted Gaussian process regression (SA-GPR)~\cite{grisafi2018} framework, which has proven to be capable of generating highly accurate models of the polarizability of organic molecules~\cite{wilkins2019} and of molecular crystals~\cite{raimbault2019}.
We trained models for the polarization $\boldsymbol{\mu}$ of pure water and for the polarizability $\boldsymbol{\alpha}$, using 1000 boxes of pure water, each containing 32 molecules, generated by replica exchange molecular dynamics (REMD) simulations with the q-TIP4P/F~\cite{habershon_competing_2009} water model~\footnote{These data first appeared in Ref.~\cite{grisafi2018} and can be found at https://archive.materialscloud.org/2018.0009/v1}.
$\boldsymbol{\mu}$ and $\boldsymbol{\alpha}$ for these systems were computed at the PBE level with Quantum ESPRESSO~\cite{giannozzi_quantum_2009}.
Full details on the training of these two models, which we will refer to as $\mu$H$_{2}$O for polarization and $\alpha$H$_{2}$O for polarizability, are found in the Supplementary Information.
$\mu$H$_{2}$O predicts the polarization per molecule with an error of $8.8\times 10^{-3}~\text{D}$, or $\sim 1\,\%$ of the intrinsic variation in the training set, and predicts an average molecular dipole moment of 2.8~D in this set, which is well within experimental error bars~\cite{gubskaya_total_2002}.
$\alpha$H$_{2}$O predicts the polarizability per molecule with an error of $5.9\times 10^{-2}~\text{a.u.}$, or $\sim 11\,\%$ of the intrinsic variation in the training set.
Fig.~\ref{fig:learning_curves} summarizes the performance of these two models in predicting the dielectric response properties of bulk water.
\begin{figure}[ht]
\centering
    \includegraphics[width=0.45\textwidth]{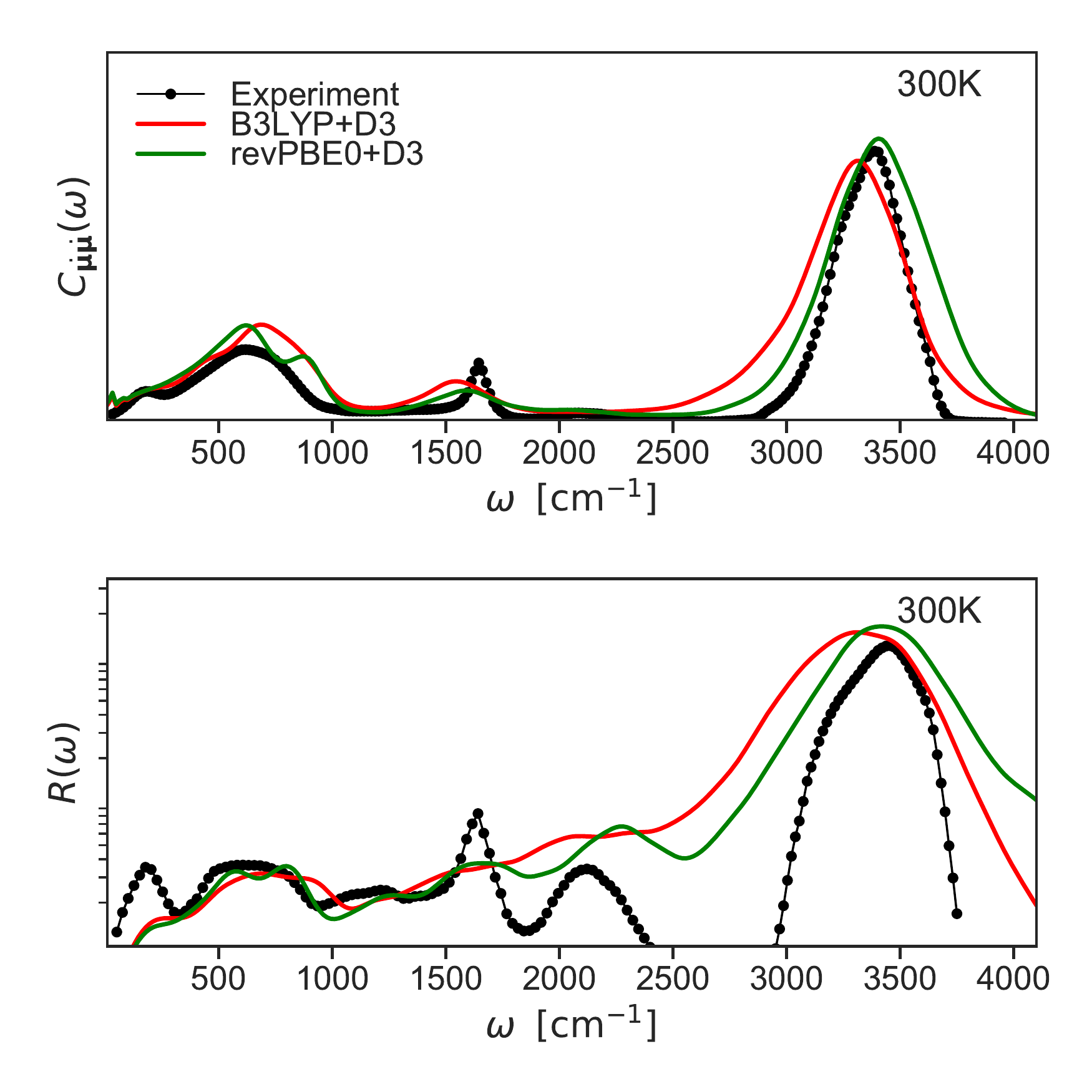}
    \includegraphics[width=0.45\textwidth]{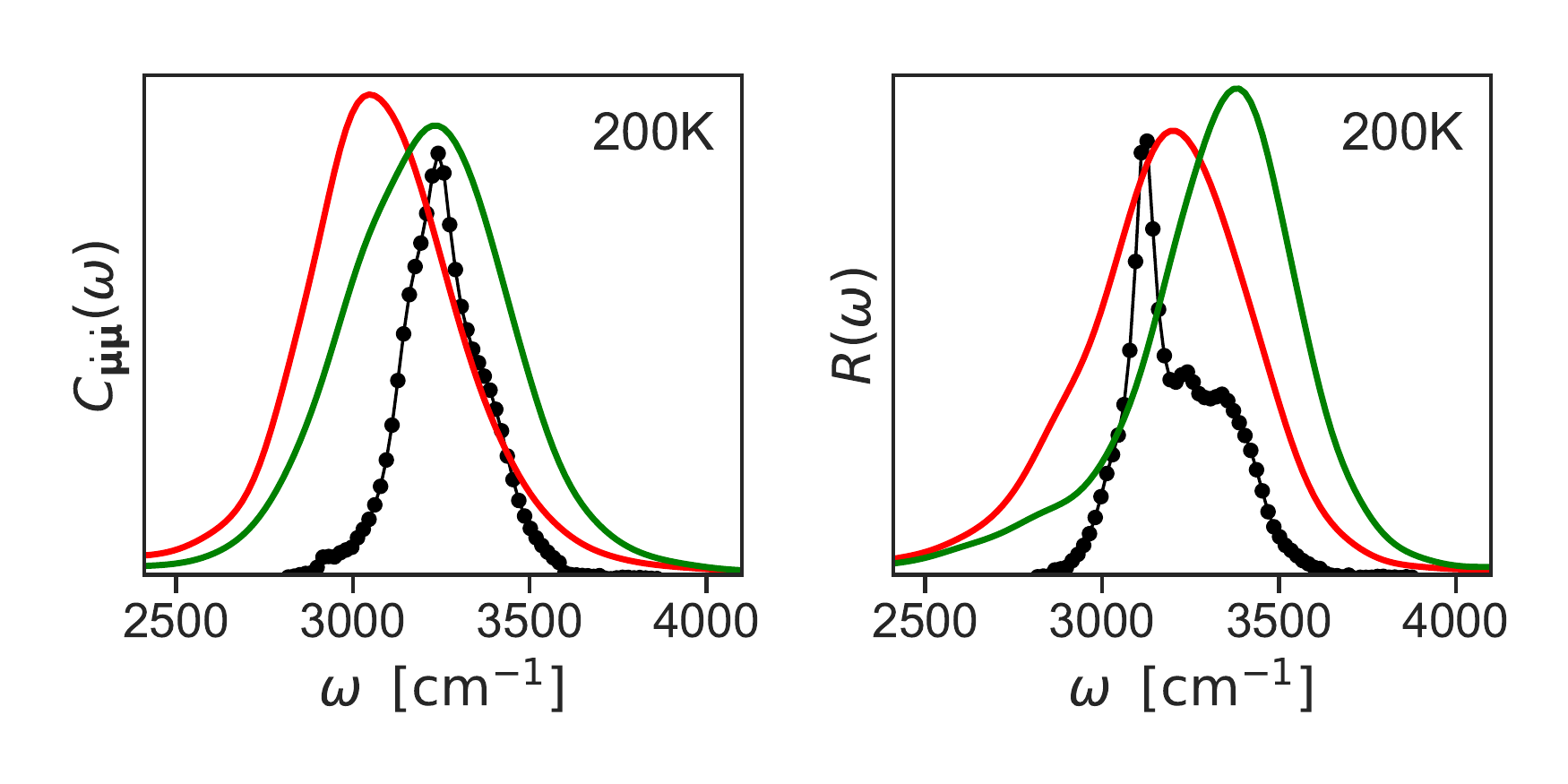}
\caption{Infra-red $C_{\dot{\bm{\mu}}\dot{\bm{\mu}}}(\omega)$ and the reduced depolarized Raman R($\omega$) spectra calculated using the autocorrelation of the time derivative of the dipole moment and the polarizabilty for hexagonal ice (bottom) at 200\,K and liquid water (top) at 300\,K using the PIGLET thermostat. Simulations were performed using neural network potentials trained on B3LYP+D3 (red) and revPBE0+D3 (green) data. Black curves show results obtained from experiments~\cite{brooker_raman_1989,bertie_infrared_1996,moberg_molecular_2017}. 
\label{fig:H2O_ML}}
\end{figure}

We first generate 100\,ps long trajectories with opt(V) PIGLET parameters, using 12 replicas for hexagonal ice at 200~K and 6 replicas for liquid water at 300\,K. 
In order to assess the comparative importance of the underlying potential energy surface and that of the approximations to quantum dynamics, we ran simulations with two ML potentials based on a Behler-Parrinello neural network framework~\cite{behl-parr07prl} -- the one introduced in Ref.~\citenum{kapil_high_2016} based on B3LYP+D3 reference calculations, and the one introduced in Ref.~\citenum{cheng_ab_2019}, based revPB0+D3. 
As shown in the SI, the peaks in the deconvoluted PIGLET velocity-velocity correlation spectra agree well with those from TRPMD, except for a red-shift of the stretching of $\approx$80~cm$^{-1}$, similar to what observed for q-TIP4P/f. For this reason, in what follows we only discuss results for PIGLET, and compare them to experiments.

As shown in the top panel of Fig.~\ref{fig:H2O_ML}, the stretching band of the IR spectrum of liquid water is well reproduced by the revPBE0+D3 potential, in agreement with Ref. \cite{marsalek_quantum_2017}, while the B3LYP+D3 potential red shifts the stretching band with respect to the experimental result~\cite{brooker_raman_1989}. The librational band is well reproduced by both potentials. The only major discrepancy in the spectra is the over-softening of the bending band, which is an artefact that was already observed in Ref.~\cite{kapil_high_2016} and linked to a shortcoming of the neural network potential. 
As shown in the middle panel of Fig.~\ref{fig:H2O_ML}, the position of the stretching band of the reduced depolarized Raman spectrum obtained from revPBE0+D3, computed using the time derivative of the instantaneous polarizability of the system following the procedure of Ref.~\cite{Liu2018}, is in excellent agreement with both experimental result and previous calculations~\cite{marsalek_quantum_2017}. As in the previous case B3LYP+D3 systematically red shifts the stretching band with respect to the experimental result~\cite{bertie_infrared_1996}.
The bending and libration portion of the spectrum, however, is reproduced only qualitatively. Due to the overwhelming intensity of the stretching band, the deconvolution algorithm introduces levels of noise in the lower-frequency region that are comparable to the intensity of the spectrum. 
At 200\,K the position of the IR stretching band of hexagonal ice calculated using the revPBE0+D3 potential is in excellent agreement with experiments~\cite{moberg_molecular_2017} as shown in the bottom panel of Fig. ~\ref{fig:H2O_ML}. 
However, the Raman stretching band of ice is blue-shifted by revPBE0+D3 and fortuitously well reproduced by the B3LYP+D3 potential,
suggesting that a significant part of the error in modelling the Raman spectrum of ice is due to the inherent limitations of DFT.
The accuracy of these results shows that the approximation inherent in applying the deconvolution procedure to the correlation spectra of non-linear dielectric responses does not introduce major artefacts, and that the level of consistency between different approximate quantum dynamics methods is equivalent or better than that between different flavors of (hybrid) density functional theory. 
The simple, robust post-processing approach we discussed here allows one to extract dynamical properties from trajectories thermostatted using PIGLET, and dramatically reduces the cost of obtaining quantum spectroscopic quantities. The implementation of the method is made available as a post-processing tool of the open-source simulation code \texttt{i-PI}~\cite{kapil_i-pi_2019}.
The accuracy of this approach is comparable with that of other, more demanding approximate methods based on path integral simulations, as we demonstrate for a harmonic oscillator (for which the method is exact), a Morse oscillator and an empirical force field for codensed phases of water.
We compared two parameterizations of the PIGLET scheme that are optimized to improve sampling efficiency and combat zero-point energy leakage, finding that the one with the weaker coupling (that we label opt(V)) makes it easier to recover the unperturbed dynamical properties. This suggests it might be worth testing parameterizations that reduce coupling strength at the expense of sampling efficiency, to achieve better agreement with TRPMD and CMD.

When combined with ML potentials trained on dispersion corrected hybrid DFT and with ML predictions of dielectric moments computed at the GGA level, the method provides IR and Raman spectra of water that are in good agreement with experiment, with a discrepancy with respect to TRPMD that is comparable to the difference between two reference electronic-structure methods.
Although the technique is derived for linear operators it also performs very well for the modelling of IR and Raman spectra, using non-linear dielectric response functions. In combination with symmetry-adapted ML models, de-convoluted PIGLET simulations open the way to accurate and computationally efficient modelling of advanced spectroscopic techniques such as second harmonic scattering and sum frequency generation.

On the whole, our work shows that the computational savings that were brought to the field by GLE techniques, and that were this far limited to static properties can be also achieved for the calculation of dynamical properties, such as  the optical spectra of condensed phase systems at low temperatures.
The reduction in computational cost will make it possible to routinely assess the importance of quantum nuclear effects in the dynamical behavior of complex condensed-phase systems, where the combination of TRPMD or CMD with first-principles energetics becomes prohibitively expensive.

\section*{Acknowledgement}
V.K. acknowledges support by the NCCR MARVEL, funded by the Swiss National Science Foundation (SNSF).
D.M.W. and M.C. acknowledge support from the European Research Council (Horizon 2020 Grant Agreement No. 677013-HBMAP).
J.L acknowledges support from a GRC Travel Grant.
This work was also supported by a grant from the Swiss National Supercomputing Centre (CSCS) under Project ID s843, and by computer time from the EPFL scientific computing centre. We thank George Trenins for sharing the results of QCMD simulations on water. We also thank Mariana Rossi and Stuart C. Althorpe for several useful discussions.

\clearpage
\ \vfill
{\centering \Huge Supporting Information\\}
\vfill
\foreach \x in {2,3,4}
{%
\clearpage


\section*{Supplementary material}

# Supporting Information

# Supporting Information

## I. MACHINE LEARNING OF DIPOLES AND POLARIZABILITIES

### A. Training Data

The $\mu$H$_2$O and $\alpha$H$_2$O models were trained using the 1000 bulk water frames generated in Ref. [1], each containing 32 molecules obtained from replica-exchange path integral molecular dynamics. The polarization, $\boldsymbol{\mu}$ and polarizability, $\boldsymbol{\alpha}$ for each frame was calculated at the DFT/PBE-USPP level using the modern theory of polarizability [2], with a plane-wave cutoff of 55 Ry and a $5 \times 5 \times 5$ $k$-point grid. This data is available on the Materials Cloud at https://archive.materialscloud.org/2018.0009/v1.

### B. Model Generation

We generated models for the polarization (a vector quantity) and the polarizability (a rank-2 symmetric tensor) using symmetry-adapted Gaussian-process regression (SA-GPR), along with $\lambda$-SOAP kernels [1].

#### 1. $\mu$H$_2$O Model

The $\mu$H$_2$O model predicts the polarization of a bulk water system $\mathcal{A}$ as a spherical tensor of order $\lambda = 1$ $\boldsymbol{\mu}(\mathcal{A})$ with components $m \in \{-1, 0, 1\}$,

$$\mu_m(\mathcal{A}) = \sum_{i \in \text{env}} \sum_{m'} k_{m,m'}^{1,\zeta}(\mathcal{A}, \mathcal{X}_i) w_{i,m'}, \tag{1}$$

where $i$ is a local environment extracted from the training set, $k_{m,m'}^{1,\zeta}(\mathcal{A}, \mathcal{X}_i)$ is a $\lambda = 1$ spherical kernel between molecule $\mathcal{A}$ and environment $\mathcal{X}_i$, and $w_{i,m'}$ is a weight; the vector of weights $\boldsymbol{w}$ is obtained by a regularized matrix overlap.

The subscript $\zeta$ in the definition of the kernel indicates that it is taken by multiplying a pure spherical kernel by a scalar kernel raised to some power,

$$k_{m,m'}^{1,\zeta}(\mathcal{A}, \mathcal{X}_i) = k_{m,m'}^{1}(\mathcal{A}, \mathcal{X}_i)(k^0(\mathcal{A}, \mathcal{X}_i))^{\zeta - 1}. \tag{2}$$

The model was optimized with respect to the following hyperparameters:

- $n_{\text{max}}$, the number of radial functions used to resolve the local density.
- $l_{\text{max}}$, the highest degree of spherical harmonic used to resolve the local density.
- $\zeta$, the exponent defined by Eq. (2).
- $n_{\text{comp}}$, the number of spherical harmonic components retained to represent each environment, selected by a farthest-point sampling (FPS) method [3].
- $n_{\text{env}}$, the number of environments drawn from the training set [4, 5] The environments to be used were chosen by FPS.
- $\sigma_G$, the width of Gaussians used to produce the SOAP density field [6].
- $r_0$ and $m$, parameters for radial scaling of the weights associated with each atom in a local environment [7]. The radial cutoff for an environment was set to $r_c = 5.0$ Å.
- $\sigma^2$, the regularization applied when inverting kernel matrices.

These hyperparameters were optimized with 5-fold cross-validation using 800 of the molecules. Table I gives the optimal hyperparameters for $\mu$H$_2$O (the same parameters were used here to build the spherical kernel and the scalar kernel used in applying Eq. (2)).

Table I. Hyperparameters used to build the $\mu$H$_2$O model, as defined in the text.

| $n_{max}$ | $l_{max}$ | $\zeta$ | $n_{comp}$ | $n_{env}$ | $\sigma_G/$Å | $r_0/$Å | $m$ | $\eta$ |
|---|---|---|---|---|---|---|---|---|
| 6 | 4 | 2 | 400 | 1000 | 0.20 | 2.0 | 5 | $1\times10^{-5}$ |

*2. $\alpha$H$_2$O Model*

The $\alpha$H$_2$O model predicts the $\lambda = 0$ (scalar) and $\lambda = 2$ components of the polarizability separately as,

$$\alpha^0(\mathcal{A}) = \bar{\alpha^0} + \sum_{i\in\text{env}} k^{0,\zeta}(\mathcal{A}, \mathcal{X}_i)w_i, \tag{3}$$

$$\alpha^2_m(\mathcal{A}) = \sum_{i\in\text{env}} \sum_{m'} k^{2,\zeta}_{m,m'}(\mathcal{A}, \mathcal{X}_i)w_{i,m'}, \tag{4}$$

where $\zeta$ is defined analogously to Eq. (2), and $m, m' \in \{-2, -1, 0, 1, 2\}$. $\bar{\alpha^0}$ is the average of the scalar part of the polarizability in the training set. The two predictions are then combined to give a prediction of the full Cartesian tensor. The definitions of these two spherical components are as given in Ref. [8]

The $\alpha$H$_2$O model is comprised of two models, which are trained separately, and each of which has its own set of hyperparameters. Though the $\lambda = 0$ kernel that is used in building the $\lambda = 2$ model does not need to be the same as that used in building the $\lambda = 0$ model, the same hyperparameters were used for each in building $\alpha$H$_2$O. Table II gives the hyperparameters used in building this model.

Table II. Hyperparameters used to build the $\alpha$H$_2$O model, as defined in the text.

| | $n_{max}$ | $l_{max}$ | $\zeta$ | $n_{comp}$ | $n_{env}$ | $\sigma_G/$Å | $r_0/$Å | $m$ | $\eta$ |
|---|---|---|---|---|---|---|---|---|---|
| $\lambda = 0$ | 8 | 6 | 2 | 400 | 1000 | 0.35 | 2.0 | 4 | $1\times10^{-6}$ |
| $\lambda = 2$ | 4 | 2 | 2 | 400 | 1000 | 0.30 | 2.0 | 4 | $1\times10^{-6}$ |

## II. VIBRATIONAL DENSITY OF STATES OF SOLID AND LIQUID WATER

Figure 1 shows the vibrational density of states computed using neural network potentials trained on revPBE0+D3 and B3LYP+D3 energetics, respectively, in combination with PIGLET and TRPMD for liquid water at 300 K and hexgonal ice at 200 K. The PIGLET simulations were performed using 12 replicas at 200 K and 6 replicas at 300 K while the TRPMD simulations required 48 replicas at 200 K and 32 replicas at 300 K. A time step of 0.25 fs was used in all the cases and the velocities and positions were sampled at each timestep for 100ps.

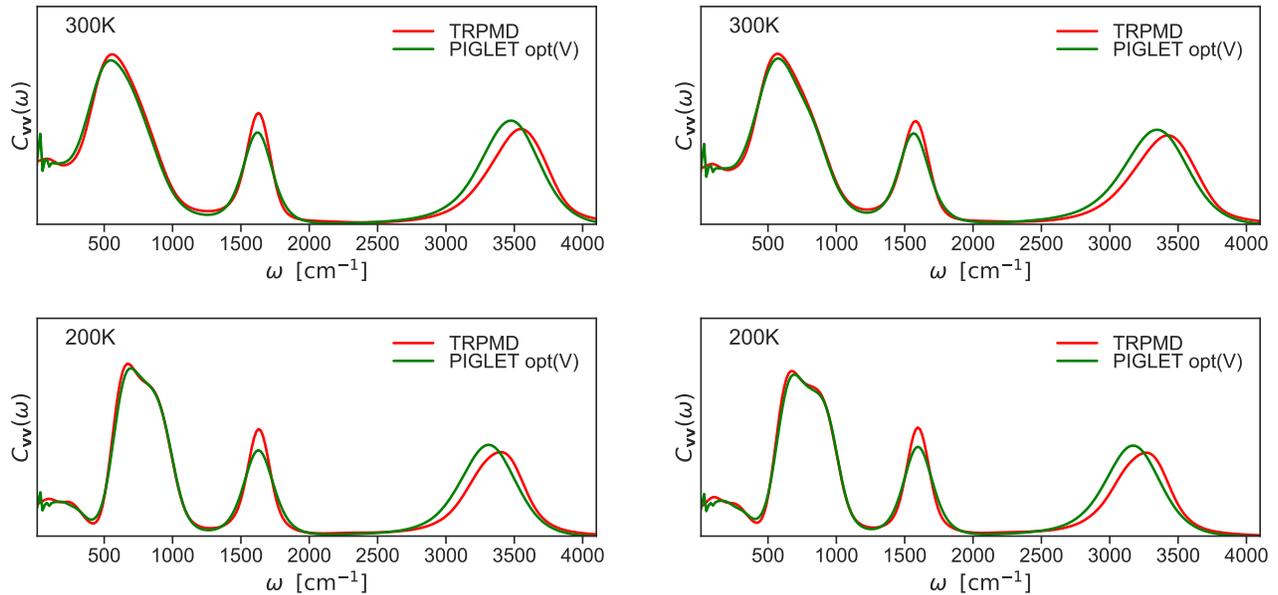

Figure 1. Vibrational density of states as calculated by the velocity autocorrelation $C_{vv}(\omega)$ for liquid water at 300 K (top) and ice (bottom) at 200 K with PIGLET (green) and TRPMD (red). The figures on the left correspond to a neural network potential trained on revPBE0+D3 energetics, those on the right to a neural network potential trained on B3LYP+D3 energetics.


} 

\end{document}